\documentclass{emulateapj}

\usepackage{graphicx}
\usepackage{amsmath}
\usepackage{natbib}
\usepackage{amssymb}

\usepackage[breaklinks,colorlinks,urlcolor=blue,citecolor=blue,linkcolor=blue]{hyperref}

\usepackage{graphicx}  
\usepackage{dcolumn}   
\usepackage{bm}        
\usepackage{amsfonts,amsmath,amssymb,mathrsfs}
\usepackage{color}

\usepackage{hyperref}
\hypersetup{
  colorlinks=true,        
  linkcolor=blue,         
  citecolor=cyan,         
}
\usepackage{natbib,times}
\definecolor{orange}{rgb}{1,0.5,0}

\newcommand{\ie}{i.e.,~}
\newcommand{\eg}{e.g.,~}

\begin{document}

\title{Using gravitational-wave observations and quasi-universal
  relations to constrain the maximum mass of neutron stars}

\author{Luciano Rezzolla\altaffilmark{1,2}, Elias
  R. Most\altaffilmark{1}, and Lukas R. Weih\altaffilmark{1}}
\altaffiltext{1}{Institut f\"ur Theoretische Physik, Max-von-Laue-Strasse 1,
  60438 Frankfurt, Germany}
\altaffiltext{2}{Frankfurt Institute for Advanced Studies,
  Ruth-Moufang-Strasse 1, 60438 Frankfurt, Germany}

\begin{abstract}
Combining the GW observations of merging systems of
binary neutron stars and quasi-universal relations, we set constraints on
the maximum mass that can be attained by nonrotating stellar models of
neutron stars. More specifically, exploiting the recent observation of
the GW event GW 170817 and drawing
from basic arguments on kilonova modeling of GRB 170817A,
together with the quasi-universal relation
between the maximum mass of nonrotating stellar models $M_{\rm TOV}$ and
the maximum mass supported through uniform rotation $M_{\rm
  max}=\left(1.20^{+0.02}_{-0.05}\right) M_{\rm TOV}$
we set limits for the maximum mass to be $ 2.01^{+0.04}_{-0.04}\leq
M_{\rm TOV}/M_{\odot}\lesssim 2.16^{+0.17}_{-0.15}$, where the lower
limit in this range comes from pulsar observations.
Our estimate, which follows a very simple line of
arguments and does not rely on the modeling of the electromagnetic
signal in terms of numerical simulations, can be further refined as new
detections become available. We briefly discuss the impact that our
conclusions have on the equation of state of nuclear matter.
\end{abstract}

\maketitle

\section{Introduction}
\label{sec:intro}

A long-awaited event took place on 2017 August 17: the Advanced
LIGO and Virgo network of GW detectors have recorded
the signal from the inspiral and merger of a binary neutron-star (BNS)
system~\citep{Abbott2017_etal}. The correlated electromagnetic signals
that have been recorded by $\sim 70$ astronomical observatories and
satellites have provided the striking confirmation that such mergers can
be associated directly with the observation of short gamma-ray bursts
(SGRBs). This event has a double significance. First, it effectively
marks the birth of multi-messenger GW astronomy. Second,
it provides important clues to solve the long-standing puzzle of the
origin of SGRBs~\citep{Eichler89, Narayan92, Rezzolla:2011,
  Berger2013b}. Numerical simulations in full general relativity of
merging BNSs have also played an important role in determining the solution of
this puzzle, and significant progress has been made over the last decade
to accurately simulate the late-inspiral, merger, and post-merger dynamics
of BNSs (see, \eg~\citet{Baiotti2016, Paschalidis2016} for recent
reviews).

Indeed, it is through the detailed analysis of the results of these
simulations that a number of recent suggestions have been made on how to
use the GW signal from merging BNSs to deduce the
properties of the system and, in particular, the equation of state (EOS)
of nuclear matter.

For instance, the changes in the phase evolution of the
GW signal during the inspiral, which depends on the tidal
deformability of stellar matter will leave a characteristic imprint on
the GW signal~\citep{Read2013, Bernuzzi2014, Hinderer2016,
  Hotokezaka2016} or in the post-merger phase. This imprint, such as the
one associated with the GW frequency at maximum amplitude 
\citep{Read2013, Bernuzzi2014, Takami2015}, can even be quasi-universal
in the sense that it depends only weakly on the EOS. Similar
considerations also apply for the post-merger signal, where the GW
spectrum exhibits characteristic frequencies~\citep{Bauswein2011,
  Takami2014}, some of which have been shown to have a quasi-universal
behavior~\citep{Bernuzzi2014, Takami2014, Takami2015, Rezzolla2016,
  Maione2017}.

Much more subtle, however, has been the task of determining the precise
fate of the binary merger product (BMP), as this depends on a number of
macroscopical factors, such as the total mass and mass ratio of the BNS
system of the angular-velocity profile~\citep{Hanauske2016}, but also of
microphysical ones, such as the efficiency of energy transport via
neutrinos~\citep{Palenzuela2015, Sekiguchi2016, Bovard2017} and the
redistribution of angular momentum via magnetic fields~\citep{Siegel2014,
  Palenzuela2015, Endrizzi2016}. While attempts have been made to
determine the mass of the binary that would lead to a prompt collapse,
\ie to a black hole within few milliseconds after merger, (see, \eg
 ~\citet{Baiotti08, Bauswein2013}), or to determine the lifetime
of the merged object (see, \eg~\citet{Lasky2013, Ravi2014, Piro2017}),
the picture on the fate of the post-merger object is still rather
uncertain. What makes such a picture complicated is the multiplicity of
stable, unstable, and metastable equilibria in which the merged object can
find itself. The importance of clarifying this picture, however, is
that understanding the ability of the merged object to sustain itself
against gravitational collapse is directly related to the maximum mass
that can be sustained against gravity, which depends on the underlying
EOS.

In this Letter, we combine the recent GW 
observation of the merging system of BNSs via the event GW 170817
\citep{Abbott2017_etal} with the existence of quasi-universal relations
regulating the equilibria of rotating and nonrotating compact stars to
set constraints on the maximum mass that can be sustained by nonrotating
stellar models of neutron stars. More specifically, after defining the
maximum mass of nonrotating models, $M_{_{\rm TOV}}$, and recalling that
the maximum mass that can be supported through uniform rotation is
$M_{\rm max}=\left(1.20^{+0.02}_{-0.02}\right)M_{\rm TOV}$ independently
of the EOS~\citep{Breu2016}, we deduce that when the merged object collapses
it has a core that is uniformly rotating and close to the maximum mass of
uniformly rotating configurations. Then our range reduces considerably
and sets the following constraint for the maximum mass
$2.01^{+0.04}_{-0.04}\leq M_{\rm TOV}/M_{\odot}\lesssim
2.16^{+0.17}_{-0.15}$. Our estimate, which is compatible with that recently
suggested by other authors~\citep{Alsing2017, Margalit2017,
  Shibata2017c, Ruiz2017}, follows a straightforward set of
considerations and does not rely on the modeling of the electromagnetic
signal via numerical-relativity simulations (as done, \eg
  by~\citet{Bovard2017} or~\cite{Shibata2017c}) but only on basic
arguments inferred from kilonova modeling \citep{Cowperthwaite2017}, can
be further refined as new observations are carried out.

\section{The basic picture}
\label{sec:basic_picture}

\begin{figure}[!t]
\includegraphics[width=1.0\columnwidth]{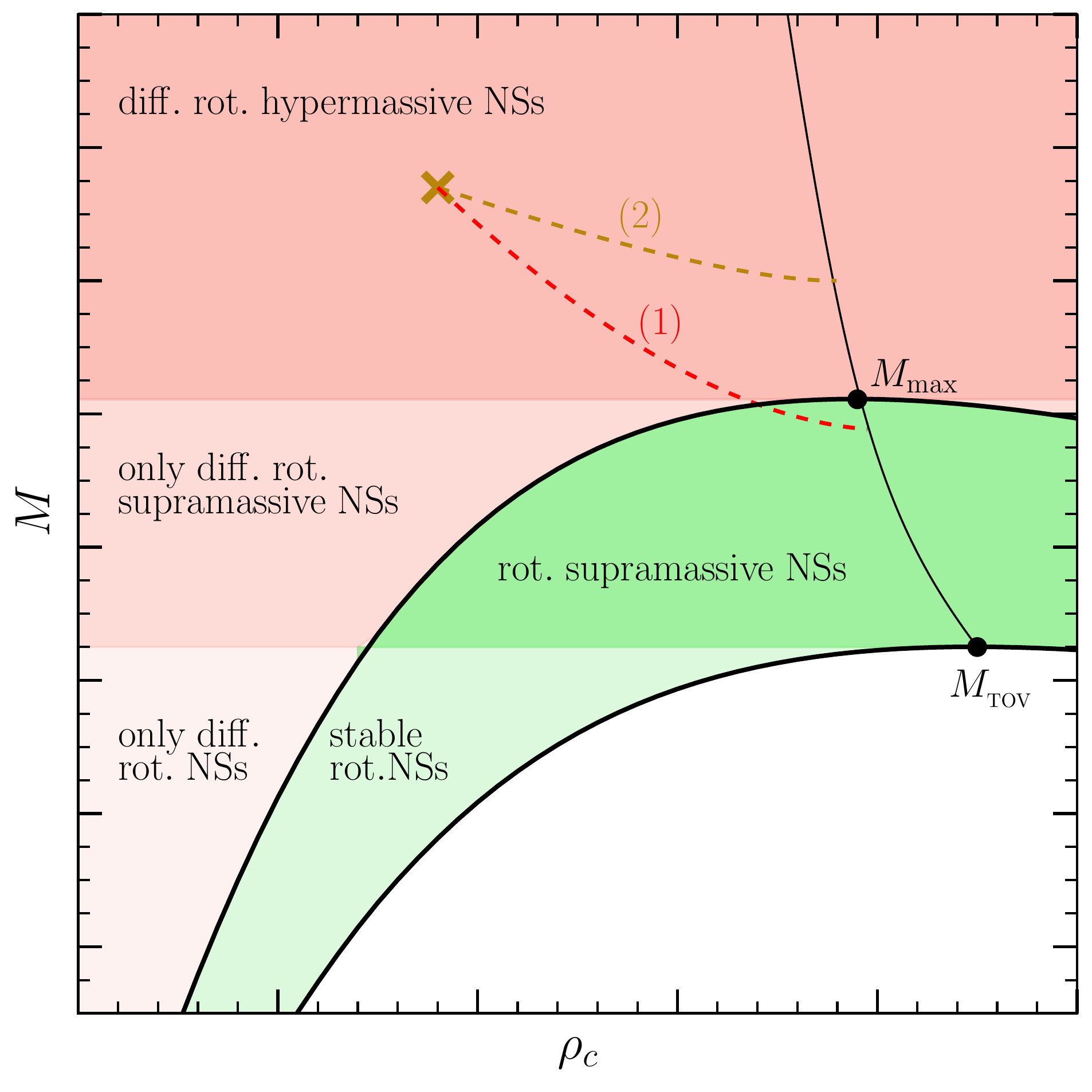}
\caption{\footnotesize Schematic diagram of the different types of
  equilibrium models for neutron stars. The golden cross marks the
  initial position of the BMP and the dashed lines its possible
  trajectories in the $(M,\rho_c)$ plane before it collapses to a black
  hole.}
\label{fig:cartoon}
\end{figure}

To illustrate the multiplicity of states that the merger of a BNS system
can lead to, we show in Fig. \ref{fig:cartoon} a schematic diagram
reporting the (gravitational) mass $M_{\rm g}$ versus the central rest-mass
density $\rho_{c}$ and thus comprising all possible stable and unstable
equilibrium states for the BMP. More specifically, shown with two solid
black lines are the sequences of nonrotating (bottom) neutron stars and
the neutron stars spinning at the mass-shedding limit. The vertical thin
black line marks the turning points of sequences with constant angular
momentum and has been shown to be a good approximation to the
neutral-stability line for uniformly as well as differentially rotating
neutron stars~\citep{Takami:2011, Weih2017}. Models on the low-density
side of this line are dynamically stable, while the ones on the
high-density side are unstable against gravitational collapse to a black
hole. Neutron stars with masses exceeding the maximum mass of
nonrotating configurations, $M_{_{\rm TOV}}$, but not the maximum mass
of uniformly rotating neutron stars, $M_{\rm max}$, are referred to as
\textit{supramassive} (SMNS), while the ones with mass higher than
$M_{\rm max}$ are called \textit{hypermassive} (HMNS; dark-red shaded
area in Fig. \ref{fig:cartoon}). The latter configurations can only be
supported by differential rotation. SMNSs, on the other hand, can be
either uniformly or differentially rotating. The uniformly rotating
models, however, are confined to the region between the nonrotating and
mass-shedding limit (green area). Outside this region, only
differentially rotating SMNSs are possible (medium-red area). Finally,
models with mass below $M_{_{\rm TOV}}$ can be rotating either
differentially (light-red area) or uniformly (light-green area).

Also reported as dashed lines are two ``trajectories'' that the BMP
produced in GW 170807 (golden cross), could have followed and that we have
indicated as ${\rm (1)}$ and ${\rm (2)}$, respectively. Both trajectories
end on the neutral-stability line because we hereafter make the working
assumption that the BMP produced in GW 170817 has indeed collapsed to a
black hole as is necessary for most models of SGRB emission from BNS
mergers, see, \eg~\citet{Rezzolla:2011, Murguia-Berthier2017}, and is also
expected to occur for most commonly used EOSs given the total mass of the
system~\citep{Abbott2017d_etal}.

In the first scenario ${\rm (1)}$, the BMP spins down and redistributes
its angular momentum due, for instance, to magnetic braking or the
development of a magnetorotational instability (see~\citet{Baiotti2016}
for a review). It does so moving on a line of almost constant baryon mass
until it eventually enters the dark-green shaded region on the stable
side of the neutral-stability line. It can then further lose
gravitational mass by spinning down until it eventually crossed the
neutral-stability line as a uniformly rotating SMNS and collapses. In the
second scenario ${\rm (2)}$, instead, the BMP passes the
neutral-stability line much more rapidly and before it can redistribute
its angular momentum, thus collapses as a differentially rotating
HMNS. This scenario, however, is unlikely when considering the
blue-kilonova signal that has been observed~\citep{Cowperthwaite2017} in
the electromagnetic counterpart of GW 170817. To produce such a signal, in
fact, ejected material with very high electron fraction $Y_e>0.25$ must
be produced, which, however, most likely originate from the hot polar
region of the BMP~\citep{Bovard2017, Metzger2017,Metzger2017b}. Hence,
the observation of such a signal inevitably requires the BMP to be
sufficiently long-lived. In particular, its lifetime should be much
longer than the timescale for reaching uniform rotation via magnetic
braking.

These considerations make the scenario $\rm (1)$ the most likely one. At
the same time, the BMP cannot have survived for very long if an SGRB was
observed only $\simeq 1\,{\rm s}$ after the merger, thus constraining the
mass of the BMP to be very close to $M_{\rm max}$ when passing the
neutral-stability line. This conclusion becomes inevitable when
considering the timescales associated with the spinning down of a
uniformly rotating neutron star. Magnetic-dipole emission, in fact, is
not sufficiently efficient and would act on much longer timescales (see,
\eg~\citet{Zhang2001}). Spin-down (and hence loss of gravitational mass)
via the GW emission driven by an ellipticity in the BMP is of course
possible, but would require unrealistic deformations to be efficient over
only $1\,{\rm s}$. We reach this conclusion by estimating the ellipticity
$\varepsilon$  required to produce such a loss by considering the typical
timescale of GW emission to be \citep{Usov1992}
\begin{align}
  \tau_\mathrm{GW}=\frac{E_\mathrm{kin}}{L_\mathrm{GW}}\,,
  \label{eqn:tau_gw}
\end{align}
where $E_\mathrm{kin}=I\Omega^2/2$ and 
\begin{align}
  L_\mathrm{GW}=\frac{32}{5}\frac{GI^2\Omega^6}{c^5}\varepsilon^2\,,
  \label{eqn:L_GW}
\end{align}
where $I$ is the moment of inertia, $\Omega$ is the rotational frequency,
$G$ is Newton's constant, and $c$ is the speed of light. Using typical values
of $I\approx 10^{45}\mathrm{g}\,\mathrm{cm}^2$, we find that
\begin{align}
  \varepsilon\gtrsim 3\times 10^{-2}\ \left(\frac{10^4\,
    \mathrm{s}^{-1}}{\Omega}\right)^{2}\left(
    \frac{1\ \mathrm{s}}{\tau_{\mathrm{GW}}}\right)^{\frac{1}{2}}\,,
  \label{eqn:tau_numerical}
\end{align}
where we have intentionally underestimated the rotational frequency
$\Omega$ of the remnant. Such high ellipticities are very unlikely even
$30\,{\rm ms}$ after the merger since the BMP becomes essentially
axisymmetric on timescales $\lesssim 50\,{\rm ms}$
\citep{Hanauske2016}. In summary, it is unlikely that the BMP has crossed
the stability line as a differentially rotating object, as this would have
happened on a timescale of tens of milliseconds. At the same time, it
must have crossed the stable region for uniform rotation very rapidly, or
it would have survived for timescales of the order of thousands of
seconds. Hence, we conclude that it must have collapsed very close if not
at the mass-shedding $M_{\rm max}$, which is what we will assume
hereafter.

\section{Quasi-universal relations}
\label{sec:quasi_universal}

A way to exploit the information from GW observations to set constraints
on the maximum mass of nonrotating stellar configurations (and hence on
the EOS) has recently been suggested by the work of~\citet{Breu2016}. In
that study, and inspired by the findings of~\citet{Yagi2013a}, it was
proposed that universal relations can be valid also away from regions of
the space of stable solutions.

In particular,~\citet{Breu2016} have shown that a universal relation is
exhibited also by equilibrium solutions of rotating relativistic stars
that are not stable. For this, uniformly rotating configurations on the
turning-point line, \ie whose mass is an extremum along a sequence of
constant angular momentum, have been considered. Such configurations are
unstable since they are found at larger central rest-mass densities than
those on the neutral-stability line and are therefore marginally
stable~\citep{Takami:2011}. In this way, it was possible to show that
this relation holds not only for the maximum value of the angular
momentum, but also for any rotation rate. The importance of this
universal relation is that it allows one to compute the maximum mass
sustainable through rapid uniform rotation, finding that, for any EOS, it
is about 20\% larger than the maximum mass supported by the corresponding
nonrotating configuration, \ie $M_{\rm max}\simeq
\left(1.20^{+0.02}_{-0.02}\right)M_{_{\rm TOV}}$, for all the EOSs
considered. The existence of such a universal relation has been confirmed
by several other authors and shown to apply also for other theories of
gravity, see \eg \cite{Staykov2016,Minamitsuji2016,Yagi2017}.

\begin{figure}[!t]
\includegraphics[width=1.0\columnwidth]{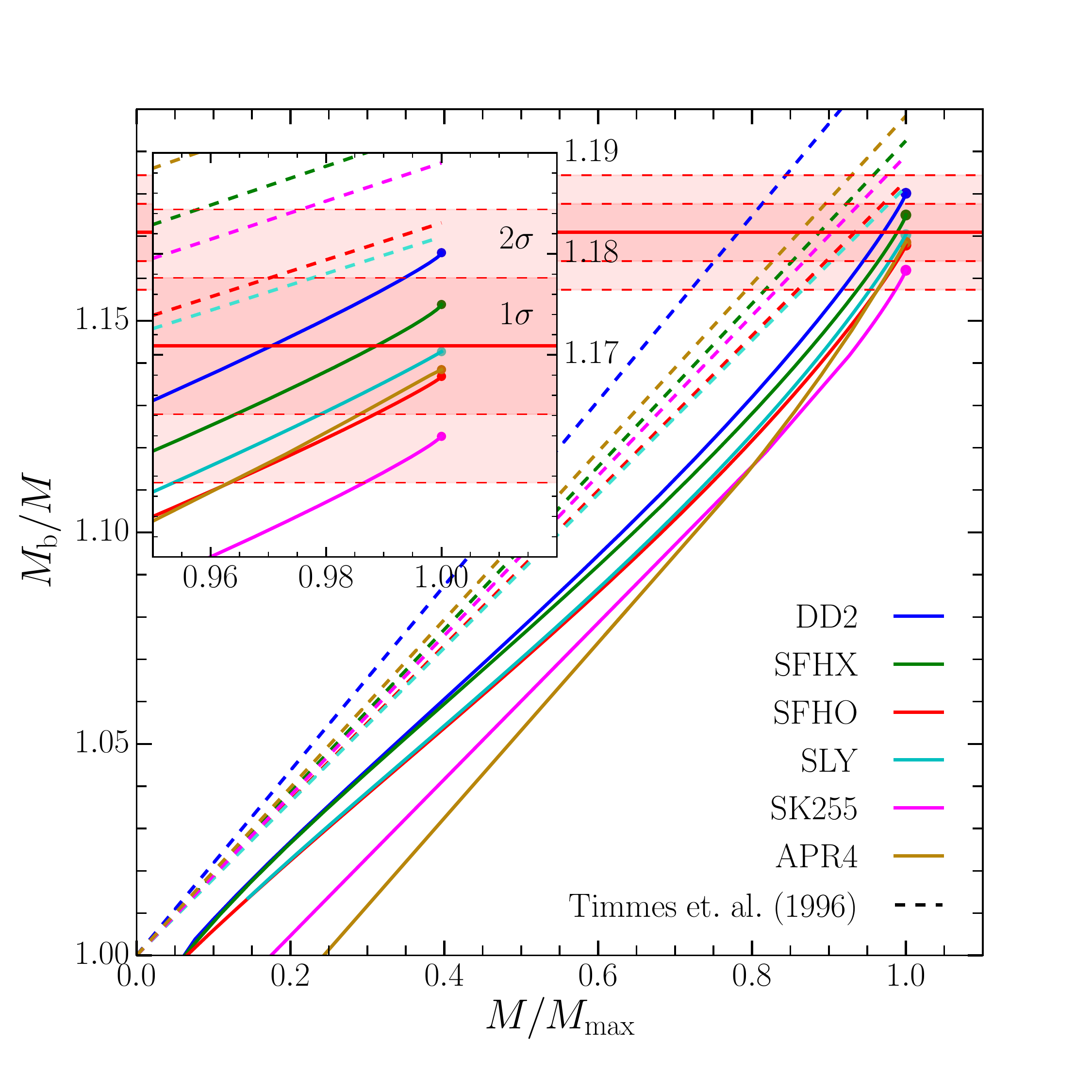}
\caption{\footnotesize Conversion factor between baryon and gravitational
  mass $M_{\rm b}/M$ of uniformly rotating configurations at the
  mass-shedding limit shown as a function of the normalized gravitational
  mass at the mass-shedding limit for different EOSs. The points of
  maximum mass are marked with dots. The red shaded areas show the
  $1\sigma$ and $2\sigma$ intervals and the horizontal red line marks the
  mean value of $M_0/M$ for the configuration with maximum mass,
  $\eta=1.171$. Also shown is the comparison to the relation derived
  in~\citet{Timmes1996} as dashed lines; note that such a relation
  overestimates the baryon mass.}
\label{fig:universal_eta}
\end{figure}

Additionally, we show a quasi-universal relation for the conversion
between gravitational mass and baryon mass, $M_{\rm b}$. In Fig.
\ref{fig:universal_eta} the conversion factor $M_{\rm b}/M$ is shown for
the sequence of uniformly rotating neutron stars spinning at the mass-shedding
limit. Interestingly, the value for the configuration with maximum mass
is only weakly dependent on the underlying EOS, and we find
\begin{equation}
\eta:=\frac{M_{\rm b}}{M_{\rm max}}\simeq 1.171\,,
\label{eq:eta}
\end{equation}
with a standard deviation of $\sigma=6.8\times 10^{-3}$. A similar
universal relation for the conversion between baryon and gravitational
mass has been proposed in~\citet{Timmes1996} and~\citet{Breu2016}. We
have used the estimate \eqref{eq:eta} here, as it refers specifically to
models that are on the mass-shedding limit and have the maximum
mass. This is presently the most accurate estimate possible for $M_{\rm
  b}/M$ at the mass-shedding limit and represents a considerable
improvement over the relation derived by~\citet{Timmes1996}, \ie $M_{\rm
  b}/M =1+0.075\,M$, which is shown as dashed lines in Fig.
\ref{fig:universal_eta}. We note that this relation is often employed,
\eg by~\citet{Piro2017}, but it systematically overestimates the relation
between the two masses by $10-25\%$.

\section{Maximum-mass constraints}

We can now use these universal relations to derive a simple mass
constraint on the EOS, making just very basic assumptions on the mass
distribution of the remnant.

As a simple parametrization, we assume that the system can be described by
the amount of ejected baryon mass $M_\mathrm{ej}$ from the inner core of
the remnant, the initial baryon mass $M_\mathrm{b}$ of the merger remnant
and the baryon mass in the uniformly rotating core $M_\mathrm{core}=\xi 
M_\mathrm{b}$. Now we can invoke simple baryon mass conservation
to concluded that $M_\mathrm{core}\left(t=0\right)=
   M_\mathrm{core}\left(t\right)+M_\mathrm{ej}$.
As we have detailed in the previous sections, we
assume that the remnant attains uniform rotation in the vicinity of the
Keplerian limit, \eg $M_\mathrm{core}\equiv M_\mathrm{core}\left(
t_{\mathrm{collapse}}\right)=M_\mathrm{b,max}$, where
$M_\mathrm{b,max}=\eta M_\mathrm{max}$ is the baryon mass at the
mass-shedding limit. Making the simplifying assumption and solving for 
$M_\mathrm{max}$ we find
\begin{align}
  M_\mathrm{max}=\eta^{-1}\left(\xi M_\mathrm{b}- \
  M_\mathrm{ej}\right)\,.
  \label{eqn:Mmax}
\end{align}
Combining this with the result from~\cite{Breu2016} we infer
\begin{align}
  M_\mathrm{TOV}=\chi^{-1}M_\mathrm{max}=\chi^{-1}\left(\xi
  M_\mathrm{g}-\eta^{-1}\ M_\mathrm{ej}\right)\,,
  \label{eqn:MTOV}
\end{align}
where $\chi=1.20^{+0.02}_{-0.02}$~\citep{Breu2016} and
$M_\mathrm{g}=\eta^{-1} M_{\mathrm{b}}=2.74^{+0.04}_{-0.01}$, which is
consistent with low-spin priors~\citep{Abbott2017_etal}.

The assumption that the core collapses exactly at the maximum
mass-shedding limit, \ie $\chi\simeq 1.2$, brings in an error that needs
to be accounted for, by considering a lower value for $\chi$ (Equation (12)
  in~\cite{Breu2016}). We thus set the lower bound to $\chi=1.15$,
corresponding to a star close to, but not at the maximum mass-shedding
limit.

\citet{Hanauske2016} have found that the mass fraction of the core after
dynamical mass ejection is roughly $\xi=0.95^{+0.06}_{-0.06}$ [see table
  II in~\cite{Hanauske2016}].  The mass of the ejecta from the core is
harder to estimate but, using standard kilonova
models~\citep{Metzger2017b, Shibata2017c}, it is reasonable to associate
them with the blue ejecta
$M^{\mathrm{blue}}_{\mathrm{ej}}=0.014^{+0.010}_{-0.010}$
\citep{Cowperthwaite2017,Drout2017}, where we have assumed a conservative
kilonova model dependent error that we use as $2\sigma$ for assigning a
Gaussian probability distribution to the blue ejecta.

\begin{figure}[!t]
\includegraphics[width=1.0\columnwidth]{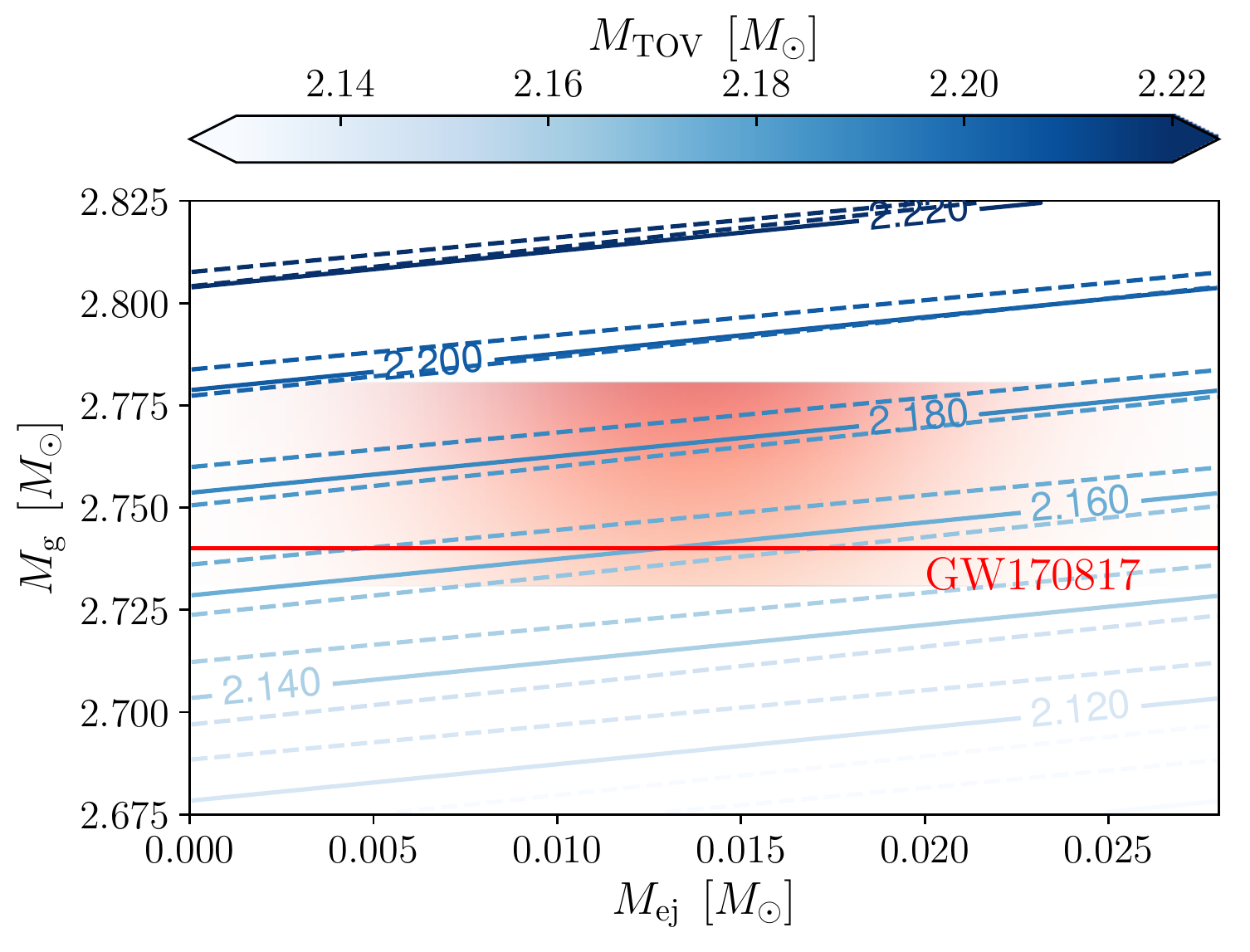}
\caption{\footnotesize Maximum-mass constraints $M_\mathrm{TOV}$ (blue
  lines) as a function of the observed gravitational mass of the BMP
  $M_{\rm g}$ and of the inferred \textit{blue} ejected mass
  $M_\mathrm{ej}$ as obtained from \eqref{eqn:MTOV}. The dashed lines
  refer to conservative error estimates of the disk mass of the merger
  product~\citep{Hanauske2016}. Shown in red is the 90\% credibility
  interval of $M_\mathrm{g}$~\citep{Abbott2017_etal}, with the red line
  denoting the most probable value from GW 170817. The transparency of
  this area reflects the probability distribution of $M_\mathrm{ej}$.}
\label{fig:mass_constraints}
\end{figure}

The resulting fit for $M_\mathrm{TOV}$ is shown in Fig.
\ref{fig:mass_constraints}, where the dashed lines refer to errors in
$\xi$ and the red shaded region is modeled with a Gaussian distribution
taking into account the errors of $M_\mathrm{ej}$. This region is framed
by the 90\% credibility levels of the binary
mass~\citep{Abbott2017_etal}.

In summary, collecting all available information, we conclude that the
maximum mass that can be supported against gravity by a compact
nonrotating star is in the range
\begin{align}
\label{eqn:MTOVmax}
2.01^{+0.04}_{+0.04}<M_{\rm TOV}/M_{\odot}<2.16^{+0.17}_{-0.15}\,,
\end{align}
where the lower limit in the range \eqref{eqn:MTOVmax} is actually
derived from accurate observations of massive pulsars in binary systems
\citep{Antoniadis2013}. 

The error corresponds to twice the standard deviation ($\sim 90\%$
confidence) computed with standard error propagation, where the
asymmetric errors in $M_\mathrm{g}$ and $\chi$ are taken into account by
computing the standard deviation for the upper and lower limit
separately. Clearly, values close to the upper and lower limits are
unlikely, given the fact that not all the values of $M_\mathrm{g}$ and
$M_\mathrm{ej}$ are equally likely (compare to the red shaded area).

Note the interesting general trend shown by the maximum mass in
Fig. \ref{fig:mass_constraints}: the estimates for $M_\mathrm{TOV}$ grow
systematically with increasingly massive binary systems and with
decreasing ejected masses (compare to the shading from light to dark blue). Hence,
future detections of merging binary systems with masses smaller than that
of GW 170817 will help set even tighter constraints on the maximum mass
$M_\mathrm{TOV}$.

\section{Conclusions}

We have combined the recent GW observations of merging
systems of binary neutron stars via the event GW 170817 with a
quasi-universal relation between the maximum mass of nonrotating stellar
models $M_{\rm TOV}$ and the maximum mass that can be supported through
uniform rotation to set new and tighter constraints on $M_{\rm TOV}$.

Our estimate follows a simple line of arguments and is based on a single
and reasonable assumption that the product of the merger measured with
GW170817 has collapsed to a rotating black hole when it had reached a
mass close to the maximum mass for SMNS models. In this way, we
can exploit quasi-universal relations to deduce that the maximum mass for
nonrotating stellar configurations should be in the range $
2.012.01^{+0.04}_{-0.04}\leq M_{\rm TOV}/M_{\odot}\lesssim
2.16^{+0.17}_{-0.15}$. We note that it is, in principle, possible to constrain
the lower limit for $M_{\rm TOV}$ also with a quasi-universal relation on
the maximum mass of a neutron star in differential rotation
\citep{Weih2017}.

A few remarks before concluding. First, a much more conservative upper
limit $M_{\rm TOV}$ can be set uniquely assuming that the maximum
nonrotating mass $M_{\rm TOV}$ cannot be smaller than the mass in the
uniformly rotating core $M_\mathrm{core}$. Taking into account the amount
of mass ejected and the conversion between baryon and gravitational mass,
this yields $M_{\rm TOV}/M_{\odot}\lesssim 2.59$. Second, our predictions
are compatible with those recently presented by~\citet{Shibata2017c,
  Margalit2017}, sharing a number of similar considerations with the
latter. However, differently from these other works, we have not employed
a simple correlation between the maximum mass-shedding mass and the
maximum nonrotating mass, or fitting formulas stemming from numerical
simulations whose error budget is uncertain~\citep{Bauswein2013}, nor
have we relied on direct comparisons with numerical-relativity
simulations for the electromagnetic emission. Rather, using basic
arguments from kilonova modeling~\citep{Cowperthwaite2017}, we have
exploited the power of universal relations for the maximum mass that are
valid for any value of the specific angular
momentum~\citep{Breu2016}. Third, the results presented here already have
a direct impact on some of the EOSs describing matter at nuclear
densities (see, \eg~\citet{Oertel2017} for a recent review). For
instance, a popular EOS routinely employed in numerical-relativity
calculations such as the DD2 EOS~\citep{Typel2010}, violates the
constraint \eqref{eqn:MTOVmax} since it has $M_{\rm
  TOV}=2.419\,M_{\odot}$; at the same time, EOSs with hyperons, \eg
BHB$\Lambda\Phi$~\citep{Banik2014} and DD2Y~\citep{Marques2017}, have
maximum masses $\lesssim 2.1\,M_{\odot}$ and therefore seem favoured
\citep{Richers2017}. Finally, we note that the procedure outlined here
and the use of stacking techniques, as those developed in the analysis of
the GW signal of BNSs~\citep{DelPozzo2013, Agathos2015, Clark2016,
  Bose2017}, can be employed in the future as the results of new
detections become available to set new and tighter constraints on the
maximum mass. New observations, in fact, will set sharper boundaries in
the probability distributions presented in
Fig. \ref{fig:mass_constraints}, thus tightening the estimates for the
maximum mass.

\acknowledgements 

It is a pleasure to thank the referee for useful suggestions and Luke
Bovard and Enping Zhou for discussions. Support comes in part from
``NewCompStar'', COST Action MP1304; LOEWE-Program in HIC for FAIR;
European Union's Horizon 2020 Research and Innovation Programme (Grant
671698) (call FETHPC-1-2014, project ExaHyPE), the ERC Synergy Grant
``BlackHoleCam: Imaging the Event Horizon of Black Holes'' (Grant
No. 610058).

\bibliographystyle{yahapj}
\bibliography{aeireferences}

\begin{thebibliography}{57}
\providecommand\natexlab[1]{#1}
\providecommand\JournalTitle[1]{#1}

\bibitem[{Abbott {et~al.}(2017{\natexlab{a}})}]{Abbott2017d_etal}
Abbott, B.~P., {et~al.} 2017{\natexlab{a}},
  \href{http://stacks.iop.org/2041-8205/848/i=2/a=L13}{\JournalTitle{Astrophys.
  J. Lett.}, 848, L13}

\bibitem[{Abbott {et~al.}(2017{\natexlab{b}})}]{Abbott2017_etal}
---. 2017{\natexlab{b}},
  \href{http://dx.doi.org/10.1103/PhysRevLett.119.161101}{\JournalTitle{Phys.
  Rev. Lett.}, 119, 161101}

\bibitem[{{Agathos} {et~al.}(2015){Agathos}, {Meidam}, {Del Pozzo}, {Li},
  {Tompitak}, {Veitch}, {Vitale}, \& {Van Den Broeck}}]{Agathos2015}
{Agathos}, M., {Meidam}, J., {Del Pozzo}, W., {et~al.} 2015,
  \href{http://dx.doi.org/10.1103/PhysRevD.92.023012}{\JournalTitle{Phys. Rev.
  D}, 92, 023012}

\bibitem[{{Alsing} {et~al.}(2017){Alsing}, {Silva}, \& {Berti}}]{Alsing2017}
{Alsing}, J., {Silva}, H.~O., \& {Berti}, E. 2017, \JournalTitle{ArXiv
  e-prints}, \href{http://arxiv.org/abs/1709.07889}{{\sffamily arXiv:1709.07889
  [astro-ph.HE]}}

\bibitem[{{Antoniadis} {et~al.}(2013){Antoniadis}, {Freire}, {Wex}, {Tauris},
  {Lynch}, {van Kerkwijk}, {Kramer}, {Bassa}, {Dhillon}, {Driebe}, {Hessels},
  {Kaspi}, {Kondratiev}, {Langer}, {Marsh}, {McLaughlin}, {Pennucci}, {Ransom},
  {Stairs}, {van Leeuwen}, {Verbiest}, \& {Whelan}}]{Antoniadis2013}
{Antoniadis}, J., {Freire}, P.~C.~C., {Wex}, N., {et~al.} 2013,
  \href{http://dx.doi.org/10.1126/science.1233232}{\JournalTitle{Science}, 340,
  448}

\bibitem[{{Baiotti} {et~al.}(2008){Baiotti}, {Giacomazzo}, \&
  {Rezzolla}}]{Baiotti08}
{Baiotti}, L., {Giacomazzo}, B., \& {Rezzolla}, L. 2008,
  \href{http://dx.doi.org/10.1103/PhysRevD.78.084033}{\JournalTitle{Phys. Rev.
  D}, 78, 084033}

\bibitem[{Baiotti \& Rezzolla(2017)}]{Baiotti2016}
Baiotti, L., \& Rezzolla, L. 2017,
  \href{http://dx.doi.org/10.1088/1361-6633/aa67bb}{\JournalTitle{Rept. Prog.
  Phys.}, 80, 096901}

\bibitem[{{Banik} {et~al.}(2014){Banik}, {Hempel}, \&
  {Bandyopadhyay}}]{Banik2014}
{Banik}, S., {Hempel}, M., \& {Bandyopadhyay}, D. 2014,
  \href{http://dx.doi.org/10.1088/0067-0049/214/2/22}{\JournalTitle{Astrohys.
  J. Suppl.}, 214, 22}

\bibitem[{{Bauswein} {et~al.}(2013){Bauswein}, {Baumgarte}, \&
  {Janka}}]{Bauswein2013}
{Bauswein}, A., {Baumgarte}, T.~W., \& {Janka}, H.-T. 2013,
  \href{http://dx.doi.org/10.1103/PhysRevLett.111.131101}{\JournalTitle{Phys.
  Rev. Lett.}, 111, 131101}

\bibitem[{{Bauswein} \& {Janka}(2012)}]{Bauswein2011}
{Bauswein}, A., \& {Janka}, H.-T. 2012,
  \href{http://dx.doi.org/10.1103/PhysRevLett.108.011101}{\JournalTitle{Phys.
  Rev. Lett.}, 108, 011101}

\bibitem[{{Berger}(2014)}]{Berger2013b}
{Berger}, E. 2014,
  \href{http://dx.doi.org/10.1146/annurev-astro-081913-035926}{\JournalTitle{Annual
  Review of Astron. and Astrophys.}, 52, 43}

\bibitem[{{Bernuzzi} {et~al.}(2014){Bernuzzi}, {Nagar}, {Balmelli}, {Dietrich},
  \& {Ujevic}}]{Bernuzzi2014}
{Bernuzzi}, S., {Nagar}, A., {Balmelli}, S., {Dietrich}, T., \& {Ujevic}, M.
  2014,
  \href{http://dx.doi.org/10.1103/PhysRevLett.112.201101}{\JournalTitle{Phys.
  Rev. Lett.}, 112, 201101}

\bibitem[{{Bose} {et~al.}(2017){Bose}, {Chakravarti}, {Rezzolla},
  {Sathyaprakash}, \& {Takami}}]{Bose2017}
{Bose}, S., {Chakravarti}, K., {Rezzolla}, L., {Sathyaprakash}, B.~S., \&
  {Takami}, K. 2017, \JournalTitle{arXiv:1705.10850},
  \href{http://arxiv.org/abs/1705.10850}{{\sffamily arXiv:1705.10850 [gr-qc]}}

\bibitem[{{Bovard} {et~al.}(2017){Bovard}, {Martin}, {Guercilena}, {Arcones},
  {Rezzolla}, \& {Korobkin}}]{Bovard2017}
{Bovard}, L., {Martin}, D., {Guercilena}, F., {et~al.} 2017,
  \JournalTitle{Phys. Rev. D}, 96, 124005

\bibitem[{{Breu} \& {Rezzolla}(2016)}]{Breu2016}
{Breu}, C., \& {Rezzolla}, L. 2016,
  \href{http://dx.doi.org/10.1093/mnras/stw575}{\JournalTitle{Mon. Not. R.
  Astron. Soc.}, 459, 646}

\bibitem[{{Clark} {et~al.}(2016){Clark}, {Bauswein}, {Stergioulas}, \&
  {Shoemaker}}]{Clark2016}
{Clark}, J.~A., {Bauswein}, A., {Stergioulas}, N., \& {Shoemaker}, D. 2016,
  \href{http://dx.doi.org/10.1088/0264-9381/33/8/085003}{\JournalTitle{Class.
  Quantum Grav.}, 33, 085003}

\bibitem[{{Cowperthwaite} {et~al.}(2017){Cowperthwaite}, {Berger}, {Villar},
  {Metzger}, {Nicholl}, {Chornock}, {Blanchard}, {Fong}, {Margutti},
  {Soares-Santos}, {Alexander}, {Allam}, {Annis}, {Brout}, {Brown}, {Butler},
  {Chen}, {Diehl}, {Doctor}, {Drout}, {Eftekhari}, {Farr}, {Finley}, {Foley},
  {Frieman}, {Fryer}, {Garc{\'{\i}}a-Bellido}, {Gill}, {Guillochon}, {Herner},
  {Holz}, {Kasen}, {Kessler}, {Marriner}, {Matheson}, {Neilsen}, {Quataert},
  {Palmese}, {Rest}, {Sako}, {Scolnic}, {Smith}, {Tucker}, {Williams},
  {Balbinot}, {Carlin}, {Cook}, {Durret}, {Li}, {Lopes}, {Louren{\c c}o},
  {Marshall}, {Medina}, {Muir}, {Mu{\~n}oz}, {Sauseda}, {Schlegel}, {Secco},
  {Vivas}, {Wester}, {Zenteno}, {Zhang}, {Abbott}, {Banerji}, {Bechtol},
  {Benoit-L{\'e}vy}, {Bertin}, {Buckley-Geer}, {Burke}, {Capozzi}, {Carnero
  Rosell}, {Carrasco Kind}, {Castander}, {Crocce}, {Cunha}, {D'Andrea}, {da
  Costa}, {Davis}, {DePoy}, {Desai}, {Dietrich}, {Drlica-Wagner}, {Eifler},
  {Evrard}, {Fernandez}, {Flaugher}, {Fosalba}, {Gaztanaga}, {Gerdes},
  {Giannantonio}, {Goldstein}, {Gruen}, {Gruendl}, {Gutierrez}, {Honscheid},
  {Jain}, {James}, {Jeltema}, {Johnson}, {Johnson}, {Kent}, {Krause}, {Kron},
  {Kuehn}, {Nuropatkin}, {Lahav}, {Lima}, {Lin}, {Maia}, {March}, {Martini},
  {McMahon}, {Menanteau}, {Miller}, {Miquel}, {Mohr}, {Neilsen}, {Nichol},
  {Ogando}, {Plazas}, {Roe}, {Romer}, {Roodman}, {Rykoff}, {Sanchez},
  {Scarpine}, {Schindler}, {Schubnell}, {Sevilla-Noarbe}, {Smith}, {Smith},
  {Sobreira}, {Suchyta}, {Swanson}, {Tarle}, {Thomas}, {Thomas}, {Troxel},
  {Vikram}, {Walker}, {Wechsler}, {Weller}, {Yanny}, \&
  {Zuntz}}]{Cowperthwaite2017}
{Cowperthwaite}, P.~S., {Berger}, E., {Villar}, V.~A., {et~al.} 2017,
  \href{http://dx.doi.org/10.3847/2041-8213/aa8fc7}{\JournalTitle{Astrophys. J.
  Lett.}, 848, L17}

\bibitem[{{Del Pozzo} {et~al.}(2013){Del Pozzo}, {Li}, {Agathos}, {Van Den
  Broeck}, \& {Vitale}}]{DelPozzo2013}
{Del Pozzo}, W., {Li}, T.~G.~F., {Agathos}, M., {Van Den Broeck}, C., \&
  {Vitale}, S. 2013,
  \href{http://dx.doi.org/10.1103/PhysRevLett.111.071101}{\JournalTitle{Phys.
  Rev. Lett.}, 111, 071101}

\bibitem[{{Drout} {et~al.}(2017){Drout}, {Piro}, {Shappee}, {Kilpatrick},
  {Simon}, {Contreras}, {Coulter}, {Foley}, {Siebert}, {Morrell}, {Boutsia},
  {Di Mille}, {Holoien}, {Kasen}, {Kollmeier}, {Madore}, {Monson},
  {Murguia-Berthier}, {Pan}, {Prochaska}, {Ramirez-Ruiz}, {Rest}, {Adams},
  {Alatalo}, {Ba{\~n}ados}, {Baughman}, {Beers}, {Bernstein}, {Bitsakis},
  {Campillay}, {Hansen}, {Higgs}, {Ji}, {Maravelias}, {Marshall}, {Moni Bidin},
  {Prieto}, {Rasmussen}, {Rojas-Bravo}, {Strom}, {Ulloa},
  {Vargas-Gonz{\'a}lez}, {Wan}, \& {Whitten}}]{Drout2017}
{Drout}, M.~R., {Piro}, A.~L., {Shappee}, B.~J., {et~al.} 2017,
  \JournalTitle{Science, in press},
  \href{http://arxiv.org/abs/1710.05443}{{\sffamily arXiv:1710.05443
  [astro-ph.HE]}}

\bibitem[{{Eichler} {et~al.}(1989){Eichler}, {Livio}, {Piran}, \&
  {Schramm}}]{Eichler89}
{Eichler}, D., {Livio}, M., {Piran}, T., \& {Schramm}, D.~N. 1989,
  \href{http://dx.doi.org/10.1038/340126a0}{\JournalTitle{Nature}, 340, 126}

\bibitem[{{Endrizzi} {et~al.}(2016){Endrizzi}, {Ciolfi}, {Giacomazzo},
  {Kastaun}, \& {Kawamura}}]{Endrizzi2016}
{Endrizzi}, A., {Ciolfi}, R., {Giacomazzo}, B., {Kastaun}, W., \& {Kawamura},
  T. 2016,
  \href{http://dx.doi.org/10.1088/0264-9381/33/16/164001}{\JournalTitle{Classical
  and Quantum Gravity}, 33, 164001}

\bibitem[{{Hanauske} {et~al.}(2017){Hanauske}, {Takami}, {Bovard}, {Rezzolla},
  {Font}, {Galeazzi}, \& {St{\"o}cker}}]{Hanauske2016}
{Hanauske}, M., {Takami}, K., {Bovard}, L., {et~al.} 2017,
  \href{http://dx.doi.org/10.1103/PhysRevD.96.043004}{\JournalTitle{Phys. Rev.
  D}, 96, 043004}

\bibitem[{{Hinderer} {et~al.}(2016){Hinderer}, {Taracchini}, {Foucart},
  {Buonanno}, {Steinhoff}, {Duez}, {Kidder}, {Pfeiffer}, {Scheel}, {Szilagyi},
  {Hotokezaka}, {Kyutoku}, {Shibata}, \& {Carpenter}}]{Hinderer2016}
{Hinderer}, T., {Taracchini}, A., {Foucart}, F., {et~al.} 2016,
  \href{http://dx.doi.org/10.1103/PhysRevLett.116.181101}{\JournalTitle{Phys.
  Rev. Lett.}, 116, 181101}

\bibitem[{{Hotokezaka} {et~al.}(2016){Hotokezaka}, {Kyutoku}, {Sekiguchi}, \&
  {Shibata}}]{Hotokezaka2016}
{Hotokezaka}, K., {Kyutoku}, K., {Sekiguchi}, Y.-i., \& {Shibata}, M. 2016,
  \href{http://dx.doi.org/10.1103/PhysRevD.93.064082}{\JournalTitle{Phys. Rev.
  D}, 93, 064082}

\bibitem[{{Lasky} {et~al.}(2014){Lasky}, {Haskell}, {Ravi}, {Howell}, \&
  {Coward}}]{Lasky2013}
{Lasky}, P.~D., {Haskell}, B., {Ravi}, V., {Howell}, E.~J., \& {Coward}, D.~M.
  2014, \href{http://dx.doi.org/10.1103/PhysRevD.89.047302}{\JournalTitle{Phys.
  Rev. D}, 89, 047302}

\bibitem[{{Maione} {et~al.}(2017){Maione}, {De Pietri}, {Feo}, \&
  {L{\"o}ffler}}]{Maione2017}
{Maione}, F., {De Pietri}, R., {Feo}, A., \& {L{\"o}ffler}, F. 2017,
  \href{http://dx.doi.org/10.1103/PhysRevD.96.063011}{\JournalTitle{Phys. Rev.
  D}, 96, 063011}

\bibitem[{{Margalit} \& {Metzger}(2017)}]{Margalit2017}
{Margalit}, B., \& {Metzger}, B.~D. 2017,
  \href{http://dx.doi.org/10.3847/2041-8213/aa991c}{\JournalTitle{Astrophys. J.
  Lett.}, 850, L19}

\bibitem[{{Marques} {et~al.}(2017){Marques}, {Oertel}, {Hempel}, \&
  {Novak}}]{Marques2017}
{Marques}, M., {Oertel}, M., {Hempel}, M., \& {Novak}, J. 2017,
  \href{http://dx.doi.org/10.1103/PhysRevC.96.045806}{\JournalTitle{\prc}, 96,
  045806}

\bibitem[{{Metzger}(2017{\natexlab{a}})}]{Metzger2017}
{Metzger}, B.~D. 2017{\natexlab{a}},
  \href{http://dx.doi.org/10.1007/s41114-017-0006-z}{\JournalTitle{Living
  Reviews in Relativity}, 20, 3}

\bibitem[{{Metzger}(2017{\natexlab{b}})}]{Metzger2017b}
---. 2017{\natexlab{b}}, \JournalTitle{ArXiv e-prints},
  \href{http://arxiv.org/abs/1710.05931}{{\sffamily arXiv:1710.05931
  [astro-ph.HE]}}

\bibitem[{{Minamitsuji} \& {Silva}(2016)}]{Minamitsuji2016}
{Minamitsuji}, M., \& {Silva}, H.~O. 2016,
  \href{http://dx.doi.org/10.1103/PhysRevD.93.124041}{\JournalTitle{Phys. Rev.
  D}, 93, 124041}

\bibitem[{{Murguia-Berthier} {et~al.}(2017){Murguia-Berthier}, {Ramirez-Ruiz},
  {Kilpatrick}, {Foley}, {Kasen}, {Lee}, {Piro}, {Coulter}, {Drout}, {Madore},
  {Shappee}, {Pan}, {Prochaska}, {Rest}, {Rojas-Bravo}, {Siebert}, \&
  {Simon}}]{Murguia-Berthier2017}
{Murguia-Berthier}, A., {Ramirez-Ruiz}, E., {Kilpatrick}, C.~D., {et~al.} 2017,
  \href{http://dx.doi.org/10.3847/2041-8213/aa91b3}{\JournalTitle{Astrophys. J.
  Lett.}, 848, L34}

\bibitem[{{Narayan} {et~al.}(1992){Narayan}, {Paczynski}, \&
  {Piran}}]{Narayan92}
{Narayan}, R., {Paczynski}, B., \& {Piran}, T. 1992,
  \href{http://dx.doi.org/10.1086/186493}{\JournalTitle{Astrophys. J. Lett.},
  395, L83}

\bibitem[{{Oertel} {et~al.}(2017){Oertel}, {Hempel}, {Kl{\"a}hn}, \&
  {Typel}}]{Oertel2017}
{Oertel}, M., {Hempel}, M., {Kl{\"a}hn}, T., \& {Typel}, S. 2017,
  \href{http://dx.doi.org/10.1103/RevModPhys.89.015007}{\JournalTitle{Reviews
  of Modern Physics}, 89, 015007}

\bibitem[{{Palenzuela} {et~al.}(2015){Palenzuela}, {Liebling}, {Neilsen},
  {Lehner}, {Caballero}, {O'Connor}, \& {Anderson}}]{Palenzuela2015}
{Palenzuela}, C., {Liebling}, S.~L., {Neilsen}, D., {et~al.} 2015,
  \href{http://dx.doi.org/10.1103/PhysRevD.92.044045}{\JournalTitle{Phys. Rev.
  D}, 92, 044045}

\bibitem[{{Paschalidis}(2017)}]{Paschalidis2016}
{Paschalidis}, V. 2017,
  \href{http://dx.doi.org/10.1088/1361-6382/aa61ce}{\JournalTitle{Classical and
  Quantum Gravity}, 34, 084002}

\bibitem[{{Piro} {et~al.}(2017){Piro}, {Giacomazzo}, \& {Perna}}]{Piro2017}
{Piro}, A.~L., {Giacomazzo}, B., \& {Perna}, R. 2017,
  \href{http://dx.doi.org/10.3847/2041-8213/aa7f2f}{\JournalTitle{Astrophys. J.
  Lett.}, 844, L19}

\bibitem[{{Ravi} \& {Lasky}(2014)}]{Ravi2014}
{Ravi}, V., \& {Lasky}, P.~D. 2014,
  \href{http://dx.doi.org/10.1093/mnras/stu720}{\JournalTitle{Mon. Not. R.
  Astron. Soc.}, 441, 2433}

\bibitem[{{Read} {et~al.}(2013){Read}, {Baiotti}, {Creighton}, {Friedman},
  {Giacomazzo}, {Kyutoku}, {Markakis}, {Rezzolla}, {Shibata}, \&
  {Taniguchi}}]{Read2013}
{Read}, J.~S., {Baiotti}, L., {Creighton}, J.~D.~E., {et~al.} 2013,
  \href{http://dx.doi.org/10.1103/PhysRevD.88.044042}{\JournalTitle{Phys. Rev.
  D}, 88, 044042}

\bibitem[{{Rezzolla} {et~al.}(2011){Rezzolla}, {Giacomazzo}, {Baiotti},
  {Granot}, {Kouveliotou}, \& {Aloy}}]{Rezzolla:2011}
{Rezzolla}, L., {Giacomazzo}, B., {Baiotti}, L., {et~al.} 2011,
  \href{http://dx.doi.org/10.1088/2041-8205/732/1/L6}{\JournalTitle{Astrophys.
  J. Letters}, 732, L6}

\bibitem[{{Rezzolla} \& {Takami}(2016)}]{Rezzolla2016}
{Rezzolla}, L., \& {Takami}, K. 2016,
  \href{http://dx.doi.org/10.1103/PhysRevD.93.124051}{\JournalTitle{Phys. Rev.
  D}, 93, 124051}

\bibitem[{{Richers} {et~al.}(2017){Richers}, {Ott}, {Abdikamalov}, {O'Connor},
  \& {Sullivan}}]{Richers2017}
{Richers}, S., {Ott}, C.~D., {Abdikamalov}, E., {O'Connor}, E., \& {Sullivan},
  C. 2017,
  \href{http://dx.doi.org/10.1103/PhysRevD.95.063019}{\JournalTitle{Phys. Rev.
  D}, 95, 063019}

\bibitem[{{Ruiz} {et~al.}(2017){Ruiz}, {Shapiro}, \& {Tsokaros}}]{Ruiz2017}
{Ruiz}, M., {Shapiro}, S.~L., \& {Tsokaros}, A. 2017, \JournalTitle{ArXiv
  e-prints}, \href{http://arxiv.org/abs/1711.00473}{{\sffamily arXiv:1711.00473
  [astro-ph.HE]}}

\bibitem[{{Sekiguchi} {et~al.}(2016){Sekiguchi}, {Kiuchi}, {Kyutoku},
  {Shibata}, \& {Taniguchi}}]{Sekiguchi2016}
{Sekiguchi}, Y., {Kiuchi}, K., {Kyutoku}, K., {Shibata}, M., \& {Taniguchi}, K.
  2016, \href{http://dx.doi.org/10.1103/PhysRevD.93.124046}{\JournalTitle{Phys.
  Rev. D}, 93, 124046}

\bibitem[{{Shibata} {et~al.}(2017){Shibata}, {Fujibayashi}, {Hotokezaka},
  {Kiuchi}, {Kyutoku}, {Sekiguchi}, \& {Tanaka}}]{Shibata2017c}
{Shibata}, M., {Fujibayashi}, S., {Hotokezaka}, K., {et~al.} 2017,
  \JournalTitle{arXiv e-prints, 1710.07579},
  \href{http://arxiv.org/abs/1710.07579}{{\sffamily arXiv:1710.07579
  [astro-ph.HE]}}

\bibitem[{{Siegel} {et~al.}(2014){Siegel}, {Ciolfi}, \&
  {Rezzolla}}]{Siegel2014}
{Siegel}, D.~M., {Ciolfi}, R., \& {Rezzolla}, L. 2014,
  \href{http://dx.doi.org/10.1088/2041-8205/785/1/L6}{\JournalTitle{Astrophys.
  J.}, 785, L6}

\bibitem[{{Staykov} {et~al.}(2016){Staykov}, {Doneva}, \&
  {Yazadjiev}}]{Staykov2016}
{Staykov}, K.~V., {Doneva}, D.~D., \& {Yazadjiev}, S.~S. 2016,
  \href{http://dx.doi.org/10.1103/PhysRevD.93.084010}{\JournalTitle{Phys. Rev.
  D}, 93, 084010}

\bibitem[{{Takami} {et~al.}(2014){Takami}, {Rezzolla}, \&
  {Baiotti}}]{Takami2014}
{Takami}, K., {Rezzolla}, L., \& {Baiotti}, L. 2014,
  \href{http://dx.doi.org/10.1103/PhysRevLett.113.091104}{\JournalTitle{Phys.
  Rev. Lett.}, 113, 091104}

\bibitem[{{Takami} {et~al.}(2015){Takami}, {Rezzolla}, \&
  {Baiotti}}]{Takami2015}
---. 2015,
  \href{http://dx.doi.org/10.1103/PhysRevD.91.064001}{\JournalTitle{Phys. Rev.
  D}, 91, 064001}

\bibitem[{{Takami} {et~al.}(2011){Takami}, {Rezzolla}, \&
  {Yoshida}}]{Takami:2011}
{Takami}, K., {Rezzolla}, L., \& {Yoshida}, S. 2011,
  \href{http://dx.doi.org/10.1111/j.1745-3933.2011.01085.x}{\JournalTitle{Mon.
  Not. R. Astron. Soc.}, 416, L1}

\bibitem[{{Timmes} {et~al.}(1996){Timmes}, {Woosley}, \& {Weaver}}]{Timmes1996}
{Timmes}, F.~X., {Woosley}, S.~E., \& {Weaver}, T.~A. 1996,
  \href{http://dx.doi.org/10.1086/176778}{\JournalTitle{\apj}, 457, 834}

\bibitem[{{Typel} {et~al.}(2010){Typel}, {R{\"o}pke}, {Kl{\"a}hn}, {Blaschke},
  \& {Wolter}}]{Typel2010}
{Typel}, S., {R{\"o}pke}, G., {Kl{\"a}hn}, T., {Blaschke}, D., \& {Wolter},
  H.~H. 2010,
  \href{http://dx.doi.org/10.1103/PhysRevC.81.015803}{\JournalTitle{Phys. Rev.
  C}, 81, 015803}

\bibitem[{{Usov}(1992)}]{Usov1992}
{Usov}, V.~V. 1992,
  \href{http://dx.doi.org/10.1038/357472a0}{\JournalTitle{Nature}, 357, 472}

\bibitem[{{Weih} {et~al.}(2018){Weih}, {Most}, \& {Rezzolla}}]{Weih2017}
{Weih}, L.~R., {Most}, E.~R., \& {Rezzolla}, L. 2018,
  \href{http://dx.doi.org/10.1093/mnrasl/slx178}{\JournalTitle{Mon. Not. R.
  Astron. Soc.}, 473, L126}

\bibitem[{{Yagi} \& {Yunes}(2013)}]{Yagi2013a}
{Yagi}, K., \& {Yunes}, N. 2013, \JournalTitle{Science}, 341, 365

\bibitem[{{Yagi} \& {Yunes}(2017)}]{Yagi2017}
---. 2017,
  \href{http://dx.doi.org/10.1016/j.physrep.2017.03.002}{\JournalTitle{\physrep},
  681, 1}

\bibitem[{{Zhang} \& {M{\'e}sz{\'a}ros}(2001)}]{Zhang2001}
{Zhang}, B., \& {M{\'e}sz{\'a}ros}, P. 2001,
  \href{http://dx.doi.org/10.1086/320255}{\JournalTitle{Astrophys. J.}, 552,
  L35}

\end{thebibliography}

\end{document}